\documentclass[article,nojss]{jss}

\usepackage[utf8]{inputenc}
\usepackage[T1]{fontenc}
\usepackage[english]{babel}
\usepackage{amsmath}
\usepackage{amssymb}
\usepackage{mathrsfs}
\usepackage{bbm}
\usepackage{amsthm}
\usepackage{color}
\usepackage{soul}
\usepackage{diagbox}
\usepackage{ulem}
\usepackage{verbatim}
\usepackage{moreverb}
\usepackage{graphicx}
\usepackage{listings}
\usepackage{dsfont}
\usepackage{Sweave}
\usepackage{subfig}
\usepackage{multirow}

\usepackage{thumbpdf,lmodern}

\usepackage{framed}

\newcommand{\slmp}{\pkg{slm}}
\newcommand{\slmf}{\code{slm}}
\DeclareMathOperator{\rank}{rank}

\author{Emmanuel Caron\\LMJL Ecole Centrale Nantes
   \And J\'er\^ome Dedecker\\Universit\'e Paris Descartes
   \And Bertrand Michel\\LMJL Ecole Centrale Nantes}
\Plainauthor{Emmanuel Caron, J\'er\^ome Dedecker, Bertrand Michel}

\title{Linear regression with stationary errors : the \proglang{R} package \pkg{slm}}
\Plaintitle{Linear regression with stationary errors : the R package slm}
\Shorttitle{Linear regression with stationary errors}

\Abstract{This paper introduces the \proglang{R} package \slmp~which stands for Stationary Linear Models. The package contains a set of statistical procedures for linear regression in the general context where the error process is strictly stationary with short memory.
We work in the setting of~\cite{hannan1973central}, who proved the asymptotic normality of the (normalized) least squares estimators (LSE) under very mild conditions on the error process. We propose different ways to estimate the asymptotic covariance matrix of the LSE, and then  to correct the type $I$ error rates of the usual tests on the parameters (as well as confidence intervals). The procedures are evaluated through different sets of simulations, and two examples of real datasets are studied.}

\Plainkeywords{Linear model, Least Squares Estimators, Stationary processes, AutoRegressive processes, Spectral density, Hypothesis testing}

\Address{
  Emmanuel Caron, Bertrand Michel\\
  Laboratoire de Math\'{e}matiques Jean Leray UMR 6629\\
  Ecole Centrale Nantes\\ 
  44300 Nantes, France\\
  E-mail: \email{emmanuel.caron@ec-nantes.fr}, \email{bertrand.michel@ec-nantes.fr}\\
  URL: \url{http://ecaron.perso.math.cnrs.fr}, \url{http://bertrand.michel.perso.math.cnrs.fr}\\
  \emph{}\\
  J\'{e}r\^{o}me Dedecker\\
  Laboratoire MAP5 UMR 8145\\
  Universit\'{e} Paris Descartes\\ 
  75006 Paris, France\\
  E-mail: \email{jerome.dedecker@parisdescartes.fr}\\
  URL: \url{http://w3.mi.parisdescartes.fr/~jdedecke/}\\
  \emph{}\\
}

\begin{document}
\Sconcordance{concordance:article_arxiv_version.tex:/Users/emmanuel/Documents/Universite-Recherche/Recherche/Articles/Article_3_slm_package/hal_arxiv_version/article_arxiv_version.Rnw:%
1 43 1 1 5 167 1 1 2 4 0 1 2 7 1 1 4 1 2 1 0 5 1 3 0 1 2 17 1 %
1 3 5 0 1 2 4 1 1 2 24 0 1 2 25 1 1 3 5 0 1 2 11 1 1 2 4 0 1 2 %
3 1 1 2 10 0 1 2 3 1 1 3 5 0 2 2 7 0 1 2 29 1 1 3 5 0 1 2 13 1 %
1 4 6 0 1 2 8 1 1 2 7 0 1 2 10 1 1 3 5 0 1 2 3 1 1 2 4 0 1 2 %
43 1 1 3 5 0 1 2 14 1 1 3 5 0 2 2 7 0 1 2 10 1 1 3 5 0 1 2 8 1 %
1 2 1 0 2 1 3 0 1 2 1 1 1 2 1 0 3 1 3 0 1 2 294 1 1 2 4 0 1 2 %
15 1 1 2 1 0 1 1 1 3 5 0 2 2 35 0 1 2 1 3 5 0 2 2 33 0 1 2 31 %
1 1 5 7 0 2 2 33 0 1 2 1 4 6 0 2 2 32 0 1 2 21 1 1 3 2 0 3 1 %
17 0 1 2 1 1 1 2 1 0 1 1 30 0 2 2 1 0 2 1 29 0 1 2 19 1 1 3 2 %
0 1 1 30 0 2 2 1 0 1 2 1 0 1 1 26 0 1 2 9 1}

\Sconcordance{concordance:article_arxiv_version.tex:/Users/emmanuel/Documents/Universite-Recherche/Recherche/Articles/Article_3_slm_package/hal_arxiv_version/article_arxiv_version.Rnw:%
1 43 1 1 5 167 1 1 2 4 0 1 2 7 1 1 4 1 2 1 0 5 1 3 0 1 2 17 1 %
1 3 5 0 1 2 4 1 1 2 24 0 1 2 25 1 1 3 5 0 1 2 11 1 1 2 4 0 1 2 %
3 1 1 2 10 0 1 2 3 1 1 3 5 0 2 2 7 0 1 2 29 1 1 3 5 0 1 2 13 1 %
1 4 6 0 1 2 8 1 1 2 7 0 1 2 10 1 1 3 5 0 1 2 3 1 1 2 4 0 1 2 %
43 1 1 3 5 0 1 2 14 1 1 3 5 0 2 2 7 0 1 2 10 1 1 3 5 0 1 2 8 1 %
1 2 1 0 2 1 3 0 1 2 1 1 1 2 1 0 3 1 3 0 1 2 294 1 1 2 4 0 1 2 %
15 1 1 2 1 0 1 1 1 3 5 0 2 2 35 0 1 2 1 3 5 0 2 2 33 0 1 2 31 %
1 1 5 7 0 2 2 33 0 1 2 1 4 6 0 2 2 32 0 1 2 21 1 1 3 2 0 3 1 %
17 0 1 2 1 1 1 2 1 0 1 1 30 0 2 2 1 0 2 1 29 0 1 2 19 1 1 3 2 %
0 1 1 30 0 2 2 1 0 1 2 1 0 1 1 26 0 1 2 9 1}


\section{Introduction}

We consider the usual linear regression model
$$
Y = X \beta + \varepsilon \, ,
$$
where $Y$ is the $n$-dimensional vector of observations, $X$ is a (possibly random) $n \times  p$ design matrix, $\beta$ is a $p$-dimensional vector of parameters, and $\varepsilon= (\varepsilon_i)_{1 \leq i \leq n}$ is the error process (with zero mean and independent of $X$). The standard assumptions are that the $\varepsilon_i$'s are independent and identically distributed (i.i.d.) with zero mean and finite variance.

In this paper, we propose to modify the standard statistical procedures (tests, confidence intervals, ...) of the linear model in the more general context where the $\varepsilon_i$'s are obtained from a strictly stationary process $(\varepsilon_i)_{i \in {\mathbb N}}$ with short memory. To be more precise, let  $\hat \beta$ denote the usual least squares estimator of $\beta$. Our approach is based on two papers: the paper by~\cite{hannan1973central} who proved the asymptotic normality of the least squares estimator $D(n)(\hat \beta - \beta)$ ($D(n)$ being the usual normalization) under very mild conditions on the design and on the error process; and a recent paper by~\cite{caron2019} who
showed that, under Hannan's conditions, the asymptotic covariance matrix of $D(n)(\hat \beta - \beta)$ can be consistently estimated.

Let us emphasize that Hannan's conditions on the error process are very mild and are satisfied for most of short-memory processes (see the discussion in Section $4.4$ of~\cite{caron2018}). Putting together the two above results, we can develop a general methodology for tests and confidence regions on the parameter $\beta$, which should be valid for most of short-memory processes. This is of course directly useful for time-series regression (we shall present in Section \hyperref[sub:S5DataCO2]{5} an application to the "Mona Loa" \proglang{R} data-set on CO2 concentration), but also in the more general context where the residuals of the linear model seem to be strongly correlated. More precisely, when checking the residuals of the linear model, if the autocorrelation function of the residuals shows significant correlations, and if the residuals can be suitably modeled by an ARMA process, then our methodology is likely to apply. We shall give an example of such a situation in Section \hyperref[sub:S5DataChine]{5} (Shangai pollution data-set).

\smallskip
Hence, the tools presented in the present paper can be seen from two different points of view:
\begin{itemize}
\item[-] as appropriate tools for time series regression with short memory error process.
\item[-] as a way to robustify the usual statistical procedures when the residuals are correlated.
\end{itemize}
\smallskip

Let us now describe the organisation of the paper. In Section \hyperref[sec:Section2]{2}, we recall the mathematical background, the consistent estimator of the asymptotic covariance matrix introduced in~\cite{caron2019} and the modified $Z$-statistics and $\chi$-square statistics for testing hypothesis on the parameter $\beta$.
In Section \hyperref[sec:slmpackage]{3} we present the  \slmp~package, and the different ways to estimate the asymptotic covariance matrix: by fitting an autoregressive process on the residuals (default procedure), by means of the kernel estimator described in~\cite{caron2019}  (theoretically valid) with a bootstrap method to choose the bandwidth (\cite{wu2009banding}), by using alternative choices of the bandwidth for the rectangular kernel (\cite{efromovich1998data}) and the quadratic spectral kernel (\cite{andrews1991heteroskedasticity}), by means of an adaptative estimator of the spectral density via Histograms (\cite{comte2001adaptive}). In Section \hyperref[sec:simu]{4}, we estimate the level of a $\chi$-square test for a linear model with random design, with different kinds of error processes and for different estimation procedures. In Section \hyperref[sec:S5Data]{5}, we present two different data sets "CO2 concentration", "Shangai pollution", and we compare the summary output of \slmf~with the usual summary output of \code{lm}.

\section{Linear regression with stationary errors}
\label{sec:Section2}

\subsection{Asymptotic results for the kernel estimator}
\label{sub:sec2_1}
We start this section by giving a short presentation of linear regression with stationary errors, more details can be found for instance in~\cite{caron2019}.   Let $\hat{\beta}$ be the usual least squares estimator for the unknown vector $\beta$. The aim is to provide hypothesis tests  and confidence regions for $\beta$ in the non i.i.d. context.

Let $\gamma$ be the autocovariance function of the error process $\varepsilon$: for any integers $k$ and $m$, let $\gamma(k) = \mathrm{Cov}(\varepsilon_{m}, \varepsilon_{m+k})$.
We also introduce the covariance matrix
\begin{equation*}
\Gamma_{n} := \left[ \gamma(j-l) \right]_{1 \leq j,l \leq n}.
\label{gamma_th}
\end{equation*}

\cite{hannan1973central} has shown a Central Limit Theorem for $\hat{\beta}$ when the error process is strictly stationary, under very mild conditions on the design and the error process. Let us notice that the design can be random or deterministic. We introduce the normalization matrix $D(n)$ which is a diagonal matrix with diagonal term $d_{j}(n) = \left \| X_{.,j} \right \|_{2}$ for $j$ in $\{1, \ldots, p\}$, where $X_{.,j}$ is the $j$th column of $X$.  Roughly speaking, Hannan's result says in particular that, given the design $X$, the vector $D(n)(\hat{\beta} - \beta)$ converges in distribution to a centered Gaussian distribution with covariance matrix $C$. As usual, in practice  the covariance matrix  $C$ is unknown and it has to be estimated. Hannan also showed the convergence of second order moment:\footnote{The transpose of a matrix $X$ is denoted by $X^{t}$.}
\begin{equation*}
\mathbb{E} \left( D(n) (\hat{\beta} - \beta) (\hat{\beta} - \beta)^{t} D(n)^{t} \Big| X \right) \xrightarrow[n \rightarrow \infty]{} C, \quad a.s.
\label{second_order_moment}
\end{equation*}
where
$$ \mathbb{E} \left( D(n) (\hat{\beta} - \beta) (\hat{\beta} - \beta)^{t} D(n)^{t} \Big| X \right) = D(n) (X^{t} X)^{-1} X^{t} \Gamma_{n} X (X^{t} X)^{-1} D(n) .$$
In this paper we propose a general plug-in approach: for some given estimator $\widehat{\Gamma}_{n}$ of $ \Gamma_{n}$, we introduce the plug-in estimator
\begin{equation*}
\widehat C = \widehat C (\widehat{\Gamma}_{n}) := D(n) (X^{t}X)^{-1} X^{t} \widehat{\Gamma}_{n} X (X^{t}X)^{-1} D(n),
\label{main_estimator}
\end{equation*}
and we use $\widehat C$ to standardize the usual statistics considered for the study of linear regression.

Let us illustrate this plug-in approach with a kernel estimator
which has been proposed in~\cite{caron2019}. For some $K$ and a bandwidth $h$, the kernel estimator $\widetilde{\Gamma}_{n,h}$ is defined by
\begin{equation}
\widetilde{\Gamma}_{n,h} = \left[ K \left( \frac{j-l}{h} \right) \tilde{\gamma}_{j-l} \right]_{1 \leq j,l \leq n},
\label{Gamma_tapered_star}
\end{equation}
where the residual based empirical covariance coefficients are defined for $0 \leq | k | \leq n-1$ by
\begin{equation}
\tilde{\gamma}_{k} = \frac{1}{n} \sum_{j=1}^{n-|k|} \hat{\varepsilon}_{j} \hat{\varepsilon}_{j+|k|}.
\label{empcovtilde}
\end{equation}
For a well-chosen kernel $K$ and under mild assumptions on the design and the error process, it has been proved in~\cite{caron2019} that
\begin{equation}
{\tilde C_n}^{-1/2} D(n)(\hat{\beta} - \beta)  \xrightarrow[n \rightarrow \infty]{\mathcal{L}} \mathcal{N}_p (0_{p}, I_p),
\label{Slut}
\end{equation}
for the plug-in estimator $ \tilde C_n := \widehat C (\widetilde{\Gamma}_{n,h_n} )$, for some suitable sequence of bandwidths $(h_n)$.

More generally, in this paper we say that an estimator $\widehat{\Gamma}_{n}$ of $\Gamma_{n}$ is  {\it consistent for estimating the covariance matrix $C$} if
$\widehat C (\widehat{\Gamma}_{n})$ is positive definite and if it converges in probability to $C$. Note that such a property requires assumptions on the design, see~\cite{caron2019}. If $\widehat C (\widehat{\Gamma}_{n})$ is consistent for estimating the covariance  matrix $C$, then $ {\widehat C(\widehat{\Gamma}_n)}^{-1/2} D(n)(\hat{\beta} - \beta)$ converges in distribution to a standard Gaussian vector.

To conclude this section, let us make some additional remarks. The interest of Caron's recent paper is that the consistency of the estimator $\widehat{C} (\widehat{\Gamma}_{n})$ is proved under Hannan's condition on the error process, which is known to be optimal with respect to the convergence in distribution (see for instance~\cite{dedecker2015optimality}), and which allows to deal with most short memory processes. But the natural estimator of the covariance matrix of $\hat{\beta}$ based on $\widehat{\Gamma}_{n}$ has been studied by many other authors in various contexts. For instance, let us mention the important line of research initiated by \cite{newey1986simple, newey1994automatic}, and the related papers by \cite{andrews1991heteroskedasticity, andrews1992improved} among others. In the paper by \cite{andrews1991heteroskedasticity}, the consistency of the estimator based on $\widehat{\Gamma}_{n}$ is proved under general conditions on the fourth order cumulants of the error process, and a data driven choice of the bandwidth is proposed. Note that these authors also considered the case of heteroskedastic processes. Most of these procedures, known as HAC (Heteroskedasticity and Autocorrelation Consistent) procedures, are implemented in the package \pkg{sandwich} by Zeileis, Lumley, Berger and Graham\footnote{See the CRAN website: \url{https://cran.r-project.org/web/packages/sandwich/index.html}.}, and presented in great detail in the paper by \cite{zeileis2004econometric}. We shall use an argument of the \pkg{sandwich} package, based on the data driven procedure described by \cite{andrews1991heteroskedasticity} at the end of Section~\ref{sub:kernel_method} (see also Section~\ref{sec:simu} for a comparison with other methods).  

\subsection{Tests and confidence regions}
\label{sub:tests}

We now present tests and confidence regions for arbitrary estimators $\widehat{\Gamma}_{n}$. The complete justifications are available for kernel estimators, see~\cite{caron2019}.

\paragraph{Z-Statistics.}
We introduce the following univariate statistics:
\begin{equation}
Z_{j} = \frac{d_{j}(n) \hat{\beta}_{j}}{\sqrt{\widehat C_{(j,j)}}}
\label{pseudoStudent}
\end{equation}
where $\widehat C = \widehat C (\widehat{\Gamma}_{n})$. If $\widehat{\Gamma}_{n}$ is consistent for estimating the covariance matrix $C$ and if $\beta_j = 0$, the distribution of $Z_{j}$ converges to a standard normal distribution when $n$ tends to infinity. We directly derive an asymptotic hypothesis test for testing $\beta_j = 0$ against $\beta_j \neq 0$ as well as an asymptotic confidence interval for $\beta_j$.

\paragraph{Chi-square statistics.} Let $A$ be a $n \times k$ matrix with $\rank(A) = k$.
Under \cite{hannan1973central}'s conditions,  $D(n)( A \hat{\beta} - A \beta)$ converges in distribution to a centered Gaussian distribution with covariance matrix $A C A^{t}$. If $\widehat{\Gamma}_{n}$ is consistent for estimating the covariance matrix $C$, then $ A \widehat C (\widehat{\Gamma}_{n})$ converges in probability to $ AC$. The matrix $  A \widehat C (\widehat{\Gamma}_{n}) A^{t} $ being symmetric positive definite, this yields
$$  W := (A \widehat C (\widehat{\Gamma}_{n}) )^{-1/2}  D(n)  A (\hat{\beta} -  \beta)  \xrightarrow[n \rightarrow \infty]{\mathcal{L}} \mathcal{N}_k (0_{k}, I_k) . $$
This last result provides asymptotical confidence regions for the vector $A \beta$. It also provides an asymptotic test for testing the hypothesis $H_0$ : $ A \beta =  0$ against $H_1$ : $ A \beta \neq  0$. Indeed, under the $H_{0}$-hypothesis, the distribution  of $\|W\|_2^2$ converges to a $\chi^{2}(k)$-distribution.

The test can be used to simplify a linear model by testing that several linear combinations between the parameters $\beta_j$ are zero, as we usually do for Anova and regression models. In particular, with $A = I_p$, the test corresponds to the test of overall significance.

\section{Introduction to linear regression with the slm package}
\label{sec:slmpackage}

Using the \slmp~package is very intuitive because the  arguments and the outputs of \slmf~are similar to those of the standard functions \code{lm}, \code{glm}, etc. The output of  the main function \slmf~is an object of class \slmf~, a specific class that has been defined for linear regression with stationary processes. The \slmf~class has methods \code{plot}, \code{summary}, \code{confint} and \code{predict}. Moreover, the class \slmf~inherits from the \code{lm} class and thus provides the output of the classical \code{lm} function.

\begin{Schunk}
\begin{Sinput}
R> library(slm)
\end{Sinput}
\end{Schunk}

The statistical tools available in \slmf~strongly depend on the choice of the covariance plug-in estimator $\widehat C (\widehat \Gamma_n)$ we use for estimating $C$. All the estimators $\widehat \Gamma_n$ proposed in \slmf~are residual-based estimators, but they rely on different approaches. In this section, we present the main functionality of \slmf~together with the different covariance plug-in estimators.

For illustrating the package, we simulate synthetic data according to the linear model:
\[Y_{i} = \beta_{1} + \beta_{2} (\log(i) + \sin(i) + Z_{i}) + \beta_{3} i + \varepsilon_{i},\]
where $Z$ is a gaussian autoregressive process of order $1$, and $\varepsilon$ is the Nonmixing process described in Section~\hyperref[sub:genmodels]{4.1}. We use the functions \code{generative\_model} and \code{generative\_process} respectively to simulate observations according to this regression design and with this specific stationary process.
More details on the designs and the processes available with \newline
\code{generative\_model} and \code{generative\_process} are given in Section~\hyperref[sub:genmodels]{4.1}.

\begin{Schunk}
\begin{Sinput}
R> n = 500
R> eps = generative_process(n,"Nonmixing")
R> design = generative_model(n,"mod2")
R> design_sim = cbind(rep(1,n), as.matrix(design))
R> beta_vec = c(2,0.001,0.5)
R> Y = design_sim %*% beta_vec + eps
\end{Sinput}
\end{Schunk}

\subsection{Linear regression via AR fitting on the residuals}

A large class of stationary processes with continuous spectral density can be well approximated by AR processes, see for instance Corollary 4.4.2 in \cite{brockwell1991time}. The covariance structure of an AR process having a closed form, it is thus easy to derive an approximation  $\widetilde{\Gamma}_{AR(p)}$ of $\Gamma_n$ by fitting an AR process on the residual process.

The AR-based method for estimating $C$ is the default version of \slmf.
This method proceeds in four main steps:
\begin{enumerate}
\item Fit an autoregressive process on the residual process $\hat{\varepsilon}$ ;
\item Compute the theoretical covariances of the fitted AR process ;
\item Plug the covariances in the Toeplitz matrix $\widetilde{\Gamma}_{AR(p)}$ ;
\item Compute $\widehat{C} = \widehat{C}(\widetilde{\Gamma}_{AR(p)})$.
\end{enumerate}

The \slmf~function fits a linear regression of the vector $Y$ on the design $X$ and then fits an AR process on the residual process using the \code{ar} function from the \pkg{stats} package. The output of the \slmf~function is an object of class \slmf. The order $p$ of the AR process is set in the argument \code{model\_selec}:
\begin{Schunk}
\begin{Sinput}
R> regslm = slm(Y ~ X1+X2, data = design, method_cov_st = "fitAR",
+  	model_selec = 3)
\end{Sinput}
\end{Schunk}
The estimated covariance is recorded as a vector in the attribute \code{cov\_st} of  \code{regslm}, which is an object of class~\slmf. The estimated covariance matrix can be computed by taking the Toeplitz matrix of \code{cov\_st}, using the \code{toeplitz} function.

\subsubsection*{Summary method}

As for \code{lm} objects, a summary of a \slmf~object is given by
\begin{Schunk}
\begin{Sinput}
R> summary(regslm)
\end{Sinput}
\begin{Soutput}
Call:
"slm(formula = myformula, data = data, x = x, y = y)"

Residuals:
     Min       1Q   Median       3Q      Max 
-13.9086  -3.4586   0.1646   3.5025  13.7488 

Coefficients:
            Estimate Std. Error z value Pr(>|z|)    
(Intercept) 2.936183   0.855214   3.433 0.000596 ***
X1          0.084387   0.082371   1.024 0.305613    
X2          0.492590   0.002738 179.938  < 2e-16 ***
---
Signif. codes:  0 '***' 0.001 '**' 0.01 '*' 0.05 '.' 0.1 ' ' 1

Residual standard error: 4.907
Multiple R-squared:  0.9953
chi2-statistic: 3.278e+04 on 2 DF,  p-value: < 2.2e-16
\end{Soutput}
\end{Schunk}
The coefficient table output by the summary provides the estimators of  the $\beta_j$'s, which are exactly the classical least squares estimators. The \code{z value} column provides the values of the  $Z_j$  statistics defined by \eqref{pseudoStudent}. The \code{Std.Error} column gives an estimation of the standard errors of the $\hat \beta_j$'s, which are taken equal to $\frac{\sqrt{\widehat{C}_{(j,j)}}}{d_j(n)}$. As with the \code{lm} function, the p-value column is the p-value for testing $\beta_j =0$ against $\beta_j \neq 0$. In this example, the small p-value for the second feature $X2$ is consistent with the value chosen for \code{beta\_vec} at the  beginning of the section. The \code{chi2-statistic} at the end of the summary is the $\chi^2$ statistic for testing the significance of the model (see the end of Section~\ref{sub:tests}) For this example, the p-value is very small, indeed the variable $X2$ has a significant effect on $Y$.

\subsubsection*{Plot argument and plot method}

\begin{table}[htp]
\begin{center}
\begin{tabular}{|c|c|}
\hline
\code{method\_cov\_st=} & plot \\
\hline
\code{fitAR} & ACF and PACF of the residual process \\
\code{kernel}  &  ACF of the residual process \\
\code{kernel} with \code{model\_selec = -1} & Graph of the estimated risk and of the estimated $\gamma(k)$'s \\
\code{spectralproj} & Estimated spectral density \\
\code{select} & ACF of the residuals up to the selected order \\
\code{efromovich} & ACF of the residuals up to the selected order \\
\code{hac} & No plot available \\
\hline
\end{tabular}
\end{center}
\caption{Plot output for each method given in the \code{method\_cov\_st} of~\slmf.}
\label{tab:plot}
\end{table}%

The \slmf~function has a \code{plot} argument: with \code{plot = TRUE}, the function plots a figure which depends on the method chosen for estimating the covariance matrix $C$. Table~\ref{tab:plot} summarizes the plots for each method given in the argument \code{method\_cov\_st}. With the AR fitting method, the argument \code{plot = TRUE} outputs the ACF and the PACF of the residual process. The ACF and PACF are computed with the functions \code{acf} and \code{pacf} of the \pkg{stats} package. As usual, the ACF and PACF graphs should help the user to choose an appropriate order for the AR process.

\begin{figure}[ht]
     \centering
     \subfloat[][ACF of the residual process.]{\includegraphics[width=0.49\textwidth]{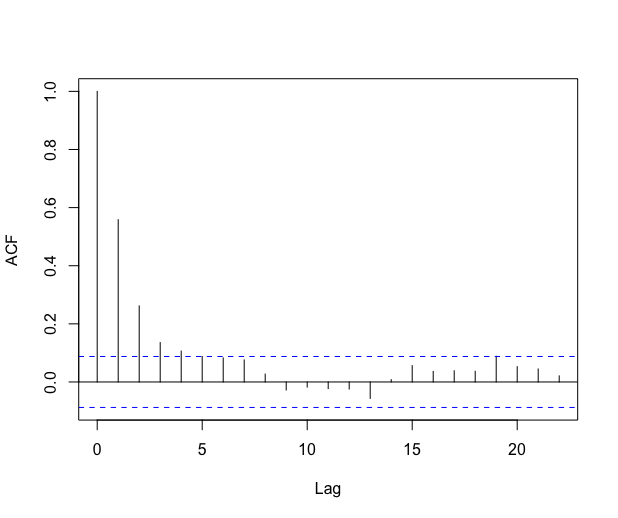}
     \label{fig:ACF_AR3}}
     \subfloat[][PACF of the residual process.]{\includegraphics[width=0.49\textwidth]{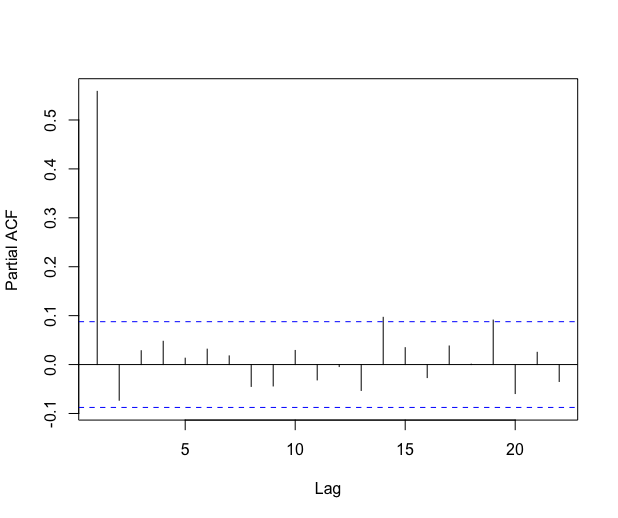}
     \label{fig:PACF_AR3}}
     \caption{Plots output by \slmf~for the fitAR method.}
     \label{fig:AR3}
\end{figure}

\begin{Schunk}
\begin{Sinput}
R> regslm = slm(Y ~ X1 + X2, data = design, method_cov_st = "fitAR",
+  	model_selec = 2, plot = TRUE)
\end{Sinput}
\end{Schunk}
The plot output by the \slmf~function for this example is given in Figure~\ref{fig:AR3}.

Since the \slmf~class inherits from the \code{lm} class, the former class comes with a \code{plot} method which is the same as for the \code{lm} class, namely the diagnostic analysis of the linear regression. The graphics are displayed using the command
\begin{Schunk}
\begin{Sinput}
R> plot(regslm)
\end{Sinput}
\end{Schunk}

\subsubsection*{Confidence intervals for the coefficients}

The \code{confint}~function computes the confidence intervals for the coefficients $\beta_j$ estimated by \slmf. These intervals are computed according to the distribution of the $Z_j$ statistics defined in \eqref{pseudoStudent}.
\begin{Schunk}
\begin{Sinput}
R> confint(regslm, level = 0.90)
\end{Sinput}
\begin{Soutput}
                    5 %      95 %
(Intercept)  1.56396552 4.3083996
X1          -0.05048587 0.2192597
X2           0.48821351 0.4969666
\end{Soutput}
\end{Schunk}

\subsubsection*{AR order selection}

The order $p$ of the AR process can be chosen at hand by setting \code{model\_selec = p}, or automatically with the AIC criterion by setting \code{model\_selec = -1}.
\begin{Schunk}
\begin{Sinput}
R> regslm = slm(Y ~ X1 + X2, data = design, method_cov_st = "fitAR",
+  	model_selec = -1)
\end{Sinput}
\end{Schunk}
The order of the fitted AR process is recorded in the \code{model\_selec}  attribute of \code{regslm}:
\begin{Schunk}
\begin{Sinput}
R> regslm@model_selec
\end{Sinput}
\begin{Soutput}
[1] 2
\end{Soutput}
\end{Schunk}
Here, the AIC criterion suggests to fit an AR(2) process on the residuals.

\subsection{Linear regression via kernel estimation of the error covariance}
\label{sub:kernel_method}

The second method for estimating the covariance matrix $C$ is the kernel estimation method \eqref{Gamma_tapered_star} studied in~\cite{caron2019}. In short, this method estimates $C$ via a smooth approximation of the covariance matrix $\Gamma_{n}$ of the residuals. This estimation of $\Gamma_{n}$ corresponds to the so-called tapered covariance matrix estimator in the literature, see for instance \cite{xiao2012covariance}, or also to the "lag-window estimator" defined in~\cite{brockwell1991time}, page $330$. It applies in particular for non negative symmetric kernels with compact support, with an integrable Fourier transform and such that $K(0) = 1$. Table~\ref{tab:kernels} gives the list of the available kernels in the package \slmp.
\begin{table}[htp]
\begin{center}
\begin{tabular}{|c|c|}
\hline
\code{kernel\_fonc =} & kernel definition \\
\hline
\code{rectangular} & $K(x) = \mathds{1}_{\{ |x| \leq 1 \}}$ \\
\hline
\code{triangle} & $K(x) = (1 - | x |) \mathds{1}_{\{ |x| \leq 1 \}}$ \\
\hline
\code{trapeze} & $K(x) = \mathds{1}_{\{ |x| \leq \delta \}} + \frac{1}{1-\delta}(1 - | x |) \mathds{1}_{\{\delta \leq |x| \leq 1 \}}$ \\
\hline
\end{tabular}
\end{center}
\caption{Available kernel functions in \slmp.}
\label{tab:kernels}
\end{table}%

It is also possible for the user to define his own kernel and to use it in the argument \code{kernel\_fonc} of the \slmf~function. Below we use the triangle kernel  which assures that the covariance matrix is positive definite.
The support of the kernel $K$ in Equation~\eqref{Gamma_tapered_star} being compact, only the terms $\tilde \gamma_{j-l}$ for small enough lag $j-l$ are kept and weighted by the  kernel in the expression of $ \widetilde{\Gamma}_{n,h}$. Rather than setting the bandwidth $h$, we select the number of $\gamma(k)$'s that should be kept (the lag) with the argument \code{model\_selec} in the \slmf~function. Then the bandwidth $h$ is calibrated accordingly, that is equal to $\texttt{model\_selec} + 1$.

\begin{figure}[h!]
\begin{center}
\includegraphics[width=0.5\textwidth]{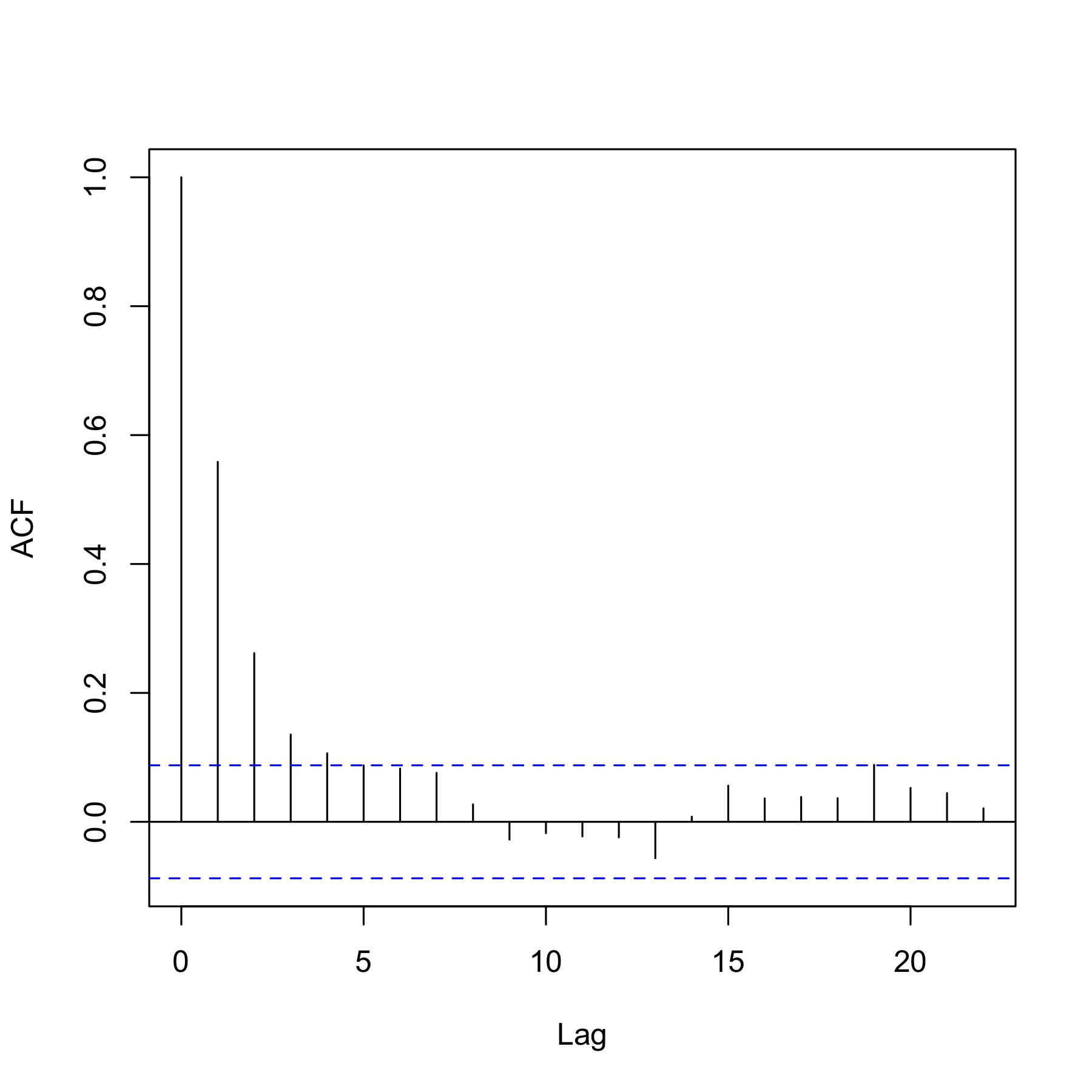}
\caption{ACF of the residual process.}
\label{fig:ACFKernel5}
\end{center}
\end{figure}

\begin{Schunk}
\begin{Sinput}
R> regslm = slm(Y ~ X1 + X2, data = design, method_cov_st = "kernel",
+  	model_selec = 5, kernel_fonc = triangle, plot = TRUE)
\end{Sinput}
\end{Schunk}
The plot output by the \slmf~function is given in Figure~\ref{fig:ACFKernel5}.

\subsubsection*{Order selection via bootstrap}

\begin{figure}[ht]
     \centering
     \subfloat[][Estimated risk error via bootstrap.]{\includegraphics[width=0.49\textwidth]{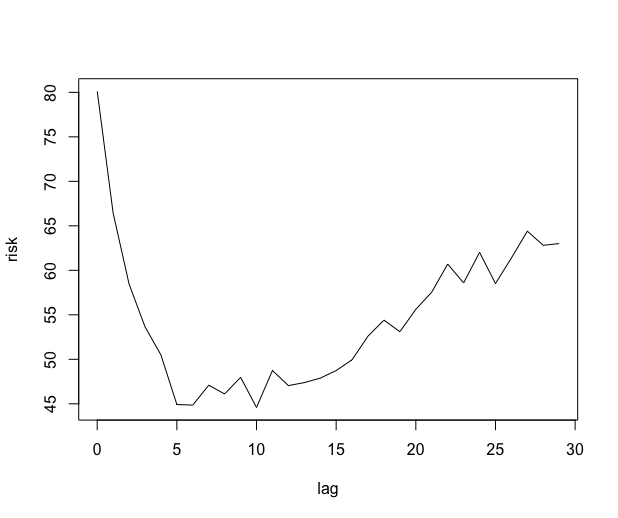}\label{fig:MSEkernelBoot}}
     \subfloat[][Estimated ACF for the selected order.]{\includegraphics[width=0.49\textwidth]{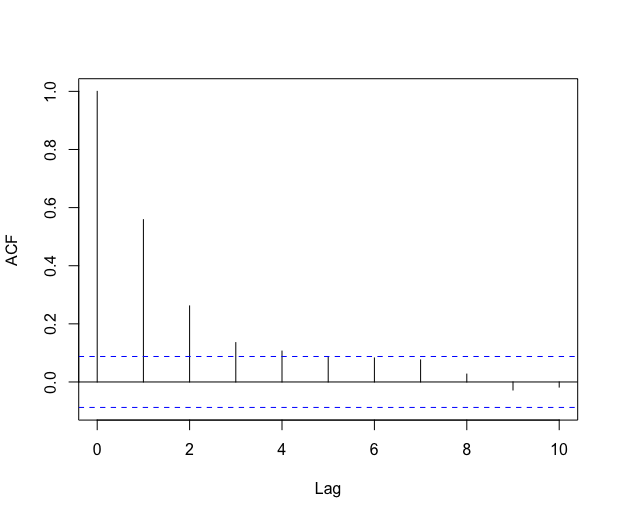}\label{fig:selectedACFkernelBoot}}
     \caption{Plots output by \slmf~for the kernel method with bootstrap selection of the order.}
     \label{fig:kernelauto}
\end{figure}

The  order  parameter can be chosen at hand as before or automatically by setting \code{model\_selec = -1}. The automatic order selection is based on the bootstrap procedure proposed by \cite{wu2009banding} for banded covariance matrix estimation. The \code{block\_size} argument sets the size of bootstrap blocks and the \code{block\_n} argument sets the number of blocks. The final order is chosen by taking the order which has the minimal risk.
Figure~\ref{fig:kernelauto} gives the plots of the estimated risk for the estimation of $\Gamma_{n}$ (left) and the final estimated ACF (right).

\begin{Schunk}
\begin{Sinput}
R> regslm = slm(Y ~ X1 + X2, data = design, method_cov_st ="kernel",
+  	model_selec = -1, kernel_fonc = triangle, model_max = 30,
+  	block_size = 100, block_n = 100, plot = TRUE)
\end{Sinput}
\end{Schunk}
The selected order is recorded in the \code{model\_selec} attribute of the \slmf~object output by the \slmf~function:
\begin{Schunk}
\begin{Sinput}
R> regslm@model_selec
\end{Sinput}
\begin{Soutput}
[1] 10
\end{Soutput}
\end{Schunk}

\subsubsection{Order selection by Efromovich's method (rectangular kernel)}

An alternative method for choosing the bandwidth in the case of the rectangular kernel has been proposed in~\cite{efromovich1998data}. For a large class of stationary processes with exponentially decaying autocovariance function (mainly the ARMA processes), Efromovich proved that the rectangular kernel is asymptotically minimax, and he proposed the following estimator:
\[\hat{f}_{J_{nr}}(\lambda) = \frac{1}{2 \pi} \sum_{k=-J_{nr}}^{k=J_{nr}} \hat{\gamma}_{k} e^{i k \lambda},\]
with the lag 
$$J_{nr} = \frac{\log(n)}{2r} \left[ 1 + (\log(n))^{-1/2} \right],$$
where $r$ is a regularity index of the autocovariance index. In practice this parameter is unknown and is estimated thanks to the algorithm proposed in the section $4$ of~\cite{efromovich1998data}.
As for the other methods, we use the residual based empirical covariances $\tilde{\gamma}_{k}$ to compute $\hat{f}_{J_{nr}}(\lambda)$.

\begin{Schunk}
\begin{Sinput}
R> regslm = slm(Y ~ X1 + X2, data = design, method_cov_st = "efromovich",
+  	model_selec = -1)
\end{Sinput}
\end{Schunk}

\subsubsection{Order Selection by Andrews's method}

Another method for choosing the bandwidth has been proposed by~\cite{andrews1991heteroskedasticity} (see the last paragraph of Section~\ref{sub:sec2_1}) and implemented in the package \pkg{sandwich} by Zeileis, Lumley, Berger and Graham (see the paper by~\cite{zeileis2004econometric}). For the \pkg{slm} package, the automatic choice of the bandwidth proposed by Andrews can be obtained  as follows 
\begin{Schunk}
\begin{Sinput}
R> regslm = slm(Y ~ X1 + X2, data = design, method_cov_st = "hac")
\end{Sinput}
\end{Schunk}
The procedure is based on the function \code{kernHAC} in the \pkg{sandwich} package. This function computes directly the covariance matrix estimator of $\hat{\beta}$, which will be recorded in the slot \code{Cov_ST} of the \code{slm} function.

Here, we take the quadratic spectral kernel 
$$
K \left( x \right) = \frac{25}{12 \pi^{2} x^{2}} \left( \frac{\sin \left( 6 \pi x / 5 \right) }{6 \pi x / 5} - \cos \left( 6 \pi x / 5 \right) \right)
$$
as suggested by Andrews (see Section $2$ in \cite{andrews1991heteroskedasticity}, or Section $3.2$ in \cite{zeileis2004econometric}), but other kernels could be used, such as Bartlett, Parzen, Tukey-Hamming, among others (see \cite{zeileis2004econometric}). 

\subsubsection*{Positive definite projection}

Depending of the method used, the matrix $\widehat{C} (\widehat{\Gamma}_{n})$ may not always be positive definite. It is the case of the kernel method with rectangular or trapeze kernel. To overcome this problem, we make the  projection of $\widehat{C} (\widehat{\Gamma}_{n})$ into the cone of positive definite matrices by applying a hard thresholding on the spectrum of this matrix: we replace all eigenvalues lower or equal to zero with the smallest positive eigenvalue of $\widehat{C} (\widehat{\Gamma}_{n})$.

Note that this projection is useless for the triangle or quadratic spectral kernels because their Fourier transform is non-negative (leading to a positive definite matrix  $\widehat{C} (\widehat{\Gamma}_{n})$). Of course, it is also useless for the \code{fitAR} and \code{spectralproj} methods. 

\subsection{Linear regression via projection spectral estimation}

The projection method relies on the ideas of \cite{comte2001adaptive}, where an adaptive nonparametric method has been  proposed for estimating the spectral density of a stationary Gaussian process.

We use the residual process as a proxy for the error process and we compute the projection coefficients with the residual-based empirical covariance coefficients $\tilde{\gamma}_{k}$, see Equation \eqref{empcovtilde}.

For some $d \in \mathbb N^*$, the estimator of the spectral density of the error process that we use is defined by computing the projection estimators for the residual process, on the basis of histogram  functions
\[\phi_{j}^{(d)} = \sqrt{\frac{d}{\pi}} \mathds{1}_{[\pi j/d, \pi (j+1)/d[}, \qquad   \ j = 0, 1, \ldots, d-1.\]
The estimator is defined by
\[\hat{f}_{d}(\lambda) = \sum_{j=0}^{d-1} \hat{a}_{j}^{(d)} \phi_{j}^{(d)},\]
where the projection coefficients are
\[\hat{a}_{j}^{(d)} = \sqrt{\frac{d}{\pi}} \left( \frac{\tilde{\gamma}_{0}}{2d} + \frac{1}{\pi} \sum_{r=1}^{n-1} \frac{\tilde{\gamma}_{r}}{r} \left[ \sin \left( \frac{\pi (j+1) r}{d} \right) - \sin \left( \frac{\pi j r}{d} \right) \right] \right).\]

\begin{figure}[ht]
\begin{center}
\includegraphics[width=0.5\textwidth]{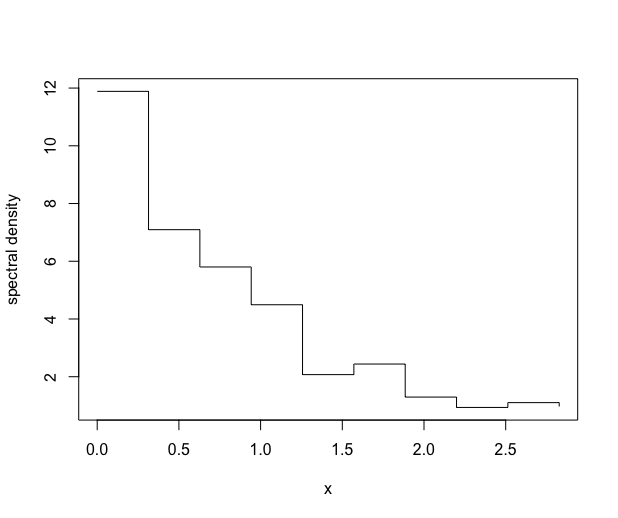}
\end{center}
\caption{Spectral density estimator by projection on the histogram basis.}
\label{fig:specdens}
\end{figure}

The Fourier coefficients of the spectral density are equal to the covariance coefficients. Thus, for $k = 1, \ldots, n-1$ it yields
\begin{eqnarray*}
\gamma_k &= & c_{k} \\
& = & \frac{2}{k} \sqrt{\frac{d}{\pi}} \sum_{j=0}^{d-1} \hat{a}_{j}^{(d)} \left[ \sin \left( \frac{k \pi (j+1)}{d} \right) - \sin \left( \frac{k \pi j}{d} \right) \right],
\end{eqnarray*}
and for $k=0$:
\begin{eqnarray*}
\gamma_0 &= c_{0} \\
& = & 2 \sqrt{\frac{\pi}{d}} \sum_{j=0}^{d-1} \hat{a}_{j}^{(d)}.
\end{eqnarray*}
This method can be proceeded in the \slmf~function by setting \code{method\_cov\_st =} \newline
\code{"spectralproj"}:
\begin{Schunk}
\begin{Sinput}
R> regslm = slm(Y ~ X1 + X2, data = design, method_cov_st = "spectralproj", 
+  	model_selec = 10, plot = TRUE)
\end{Sinput}
\end{Schunk}
The graph of the estimated spectral density can be plotted by setting \code{plot = TRUE} in the \slmf~function, see Figure~\ref{fig:specdens}.

\subsubsection*{Model selection}

The Gaussian model selection method proposed in~\cite{comte2001adaptive} follows the ideas of Birg\'e and Massart, see for instance~\cite{massart2007concentration}. It consists in minimizing the $l_2$ penalized criterion, see Section~5 in \cite{comte2001adaptive}:
$$ \mbox{crit}(d) :=  - \sum_{j=0} ^{d-1}  \left[ \hat{a}_{j}^{(d)} \right] ^2  + c \frac dn $$
where $c$ is a multiplicative constant that in practice can be calibrated using the slope heuristic method, see~\cite{birge2007minimal,baudry2012slope} and the \proglang{R} package \pkg{capushe}.
\begin{Schunk}
\begin{Sinput}
R> regslm = slm(Y ~ X1 + X2, data = design, method_cov_st = "spectralproj", 
+  	model_selec = -1, model_max = 50, plot = TRUE)
\end{Sinput}
\end{Schunk}
The selected dimension is recorded in the \code{model\_selec} attribute of the \slmf~object output by the \slmf~function:
\begin{Schunk}
\begin{Sinput}
R> regslm@model_selec
\end{Sinput}
\begin{Soutput}
[1] 8
\end{Soutput}
\end{Schunk}
The slope heuristic algorithm here selects an Histogram on a regular partition of size $8$ over the interval $[0, \pi]$ to estimate the spectral density.

\subsection{Linear regression via masked covariance estimation}

This method is a full-manual method for estimating the covariance matrix $C$ by only selecting covariance terms from the residual covariances $\tilde{\gamma}_{k}$ defined by \eqref{empcovtilde}. Let $I$ be a set of positive integers, then we  consider
$$\hat \gamma_I(k) := \tilde{\gamma}_{k} \mathbbm{1}_{k \in I\cup \{0\} }  \hskip 1cm 0 \leq |k| \leq n-1 $$
and then we define the estimated covariance marix $\widehat{\Gamma}_{I}$ by taking the Toeplitz matrix of the vector $\hat \gamma_I$. This estimator is a particular example of masked sample covariance estimator, as introduced by \cite{chen2012masked}, see also \cite{levina2012partial}. Finally we derive from $\widehat \Gamma_I$ an estimator $\widehat C (\widehat \Gamma_I) $ for $C$.

The next instruction selects the coefficients 0, 1, 2 and 4 from the residual covariance terms:
\begin{Schunk}
\begin{Sinput}
R> regslm = slm(Y ~ X1 + X2, data = design, method_cov_st = "select",
+  	model_selec = c(1,2,4))
\end{Sinput}
\end{Schunk}
The positive lags of the selected covariances are recordered in the \code{model\_selec} argument. Let us notice that the variance $\gamma_{0}$ is automatically selected.

As for the kernel method, the resulting covariance matrix may not be positive definite. If it is the case, the positive definite projection method, described at the end of the section~\ref{sub:kernel_method}, is used.

\subsection{Linear regression via manual plugged covariance matrix}

This last method is a direct plug-in method.  The user proposes his own vector estimator $\hat{\gamma}$ of $\gamma$ and then the Toeplitz matrix $\widehat \Gamma_{n}$ of the vector $\hat \gamma$ is used for estimating $C$ with $\widehat C (\widehat \Gamma_{n}) $.
\begin{Schunk}
\begin{Sinput}
R> v = rep(0,n)
R> v[1:10] = acf(epsilon, type = "covariance", lag.max = 9)$acf
R> regslm = slm(Y ~ X1 + X2, data = design, cov_st = v)
\end{Sinput}
\end{Schunk}

The user can also propose his own covariance matrix $\widehat \Gamma_{n}$ for estimating $C$.
\begin{Schunk}
\begin{Sinput}
R> v = rep(0,n)
R> v[1:10] = acf(epsilon, type = "covariance", lag.max = 9)$acf
R> V = toeplitz(v)
R> regslm = slm(Y ~ X1 + X2, data = design, Cov_ST = V)
\end{Sinput}
\end{Schunk}

Let us notice that the user must verify that the resulting covariance matrix is positive definite. The positive definite projection algorithm is not used with this method.

\section{Numerical experiments and method comparisons}
\label{sec:simu}

This section summarizes an extensive study which has been carried out to compare the performances of the different approaches presented before in the context of linear model with short range dependent stationary errors.

\subsection{Description of the generative models}
\label{sub:genmodels}

We first present the five generative models for the errors that we consider in the paper. We choose different kinds of processes to reflect the diversity of short-memory processes.

\begin{itemize}
\item {\bf AR1 process.} The AR1 process  is a gaussian AR(1)  process defined by:
\[\varepsilon_{i} - 0.7 \varepsilon_{i-1} = W_{i},\]
where $W_{i}$ is a standard gaussian distribution $\mathcal{N}(0,1)$.

\item {\bf AR12 process.} The AR12 process is a seasonal AR(12)  process defined by:
\[\varepsilon_{i} - 0.5 \varepsilon_{i-1} - 0.2 \varepsilon_{i-12} = W_{i},\]
where $W_{i}$ is a standard gaussian distribution $\mathcal{N}(0,1)$.
When studying monthly data-sets, one usually observes a seasonality of order $12$. For example, when looking at climate data (such as the "CO2 concentration" dataset of Section \ref{sec:S5Data}), the data are often collected per month, and the same phenomenon tends to repeat every year. Even if the design integrates the deterministic part of the seasonality, a correlation of order $12$ remains usually present in the residual process.

\item {\bf MA12 process.} The MA12 is also a seasonal process defined by:
\[\varepsilon_{i} = W_{i} + 0.5 W_{i-2} + 0.3 W_{i-3} + 0.2 W_{i-12},\]
where the $(W_{i})$'s are i.i.d. random variables following Student's distribution with $10$ degrees of freedom.

\item {\bf Nonmixing process.} The three processes described above are basic ARMA processes, whose innovations have absolutely continuous distributions; in particular, they are strongly mixing in the sense of~\cite{rosenblatt1956central}, with a geometric decay of the mixing coefficients (in fact the MA12 process is even $12$-dependent, which means that the mixing coefficient $\alpha(k) = 0$ if $k > 12$). Let us now describe a more complicated process: let  $(Z_{1}, \ldots, Z_{n})$ satisfying the AR(1) equation
$$Z_{i+1} = \frac{1}{2} (Z_{i} + \eta_{i+1}),$$
where $Z_{1}$ is uniformly distributed over $[0,1]$ and the $\eta_{i}$'s are i.i.d. random variables with distribution $\mathcal{B}(1/2)$, independent of $Z_{1}$.
The process $(Z_{i})_{i \geq 1}$ is a strictly stationary Markov chain, but it is not $\alpha$-mixing in the sense of Rosenblatt (see~\cite{bradley1985basic}). Let now $Q_{0,\sigma^{2}}$ be the inverse of the cumulative distribution function of a centered Gaussian distribution with variance $\sigma^{2}$ (for the simulations below, we choose $\sigma^{2} = 25$). The Nonmixing process is then defined by
\[\varepsilon_{i} = Q_{0,\sigma^{2}}(Z_{i}).\]
The sequence $(\varepsilon_{i})_{i \geq 1}$ is also a stationary Markov chain (as an invertible function of a stationary Markov chain). By construction, $\varepsilon_{i}$ is $\mathcal{N}(0, \sigma^{2})$-distributed, but the sequence $(\varepsilon_{i})_{i \geq 1}$ is not a Gaussian process (otherwise it would be mixing in the sense of Rosenblatt).
Although it is not obvious, one can prove that the process $(\varepsilon_{i})_{i \geq 1}$ satisfies Hannan's condition (see~\cite{caron2019}, Section $4.2$).

\item {\bf  Sysdyn process.}
The four processes described above have the property of "geometric decay of correlations", which means that the $\gamma(k)$'s tend to $0$ at an exponential rate. However, as already pointed out in the introduction, Hannan's condition is valid for most of short memory processes, even for processes with slow decay of correlations (provided that the $\gamma(k)$'s are summable). Hence, our last example will be a non-mixing process (in the sense of Rosenblatt), with an arithmetic decay of the correlations.

For $\gamma \in ]0,1[$,  the intermittent map $\theta_{\gamma} : [0,1] \mapsto [0,1]$ introduced in \cite{liverani1999probabilistic} is defined by
\[\theta_{\gamma}(x) =
\left\{
\begin{array}{r c l}
x(1 + 2^{\gamma} x^{\gamma}) \qquad &\text{if}& \ x \in [0, 1/2[ \\
2x - 1 \qquad &\text{if}& \ x \in [1/2, 1].\\
\end{array}
\right.\]

It follows from \cite{liverani1999probabilistic} that there exists a unique $\theta_{\gamma}$-invariant probability measure $\nu_{\gamma}$. The Sysdyn process is then defined by
$$\varepsilon_{i} = \theta_{\gamma}^{i}.$$
From~\cite{liverani1999probabilistic}, we know that, on the probability space $([0,1], \nu_{\gamma})$, the autocorrelations $\gamma(k)$ of the stationary process $(\varepsilon_{i})_{i \geq 1}$ are exactly of order $k^{-(1-\gamma)/\gamma}$. Hence $(\varepsilon_{i})_{i \geq 1}$ is a short memory process provided $\gamma \in ]0, 1/2[$. Moreover, it has been proved in Section $4.4$ of \cite{caron2018} that $(\varepsilon_{i})_{i \geq 1}$ satisfies Hannan's condition in the whole short-memory range, that is for $\gamma \in ]0, 1/2[$.
For the simulations below, we took $\gamma = 1/4$, which give autocorrelations $\gamma(k)$ of order $k^{-3}$.
\end{itemize}

\medskip

The linear regression models simulated in the experiments  all have the following form:
\begin{equation}
Y_{i} = \beta_{1} + \beta_{2} (\log(i) + \sin(i) + Z_{i}) + \beta_{3} i + \varepsilon_{i}, \hskip 1cm \textrm{for all $i$ in } \{1, \ldots, n\},
\label{reglinsimu}
\end{equation}

where $Z$ is a gaussian autoregressive process of order $1$ and $\varepsilon$ is one of the stationary processes defined above. For the simulations, $\beta_{1}$ is always equal to $3$. All the error processes presented above can be simulated with the \slmp~package with the \code{generative\_process}~function. The design can be simulated with the \code{generative\_model}~function.

\subsection{Automatic calibration of the tests}

 It is of course of first importance to provide hypothesis tests with correct significance levels or at least  with correct asymptotical significance levels, which is possible if the estimator $\widehat \Gamma_{n}$ of the covariance matrix $\Gamma_{n}$ is consistent for estimating $C$. For instance, the results of \cite{caron2019} show that it is possible to construct statistical tests with correct asymptotical significance levels.  However in practice such asymptotical results are not sufficient since they do not indicate how to tune the bandwidth on a given dataset. This situation makes the practice of linear regression with dependent errors really more difficult than linear regression with i.i.d. errors. This problem happens for several methods given before : order choice for the \code{fitAR} method, bandwidth choice for the \code{kernel} method, dimension selection for the  \code{spectralproj} method.

 It is a tricky issue to design a data driven procedure for choosing test parameters in order to have to correct Type I Error. Note that unlike with supervised problems and density estimation, it is not possible to calibrate hypothesis tests in practice using cross validation approaches. We thus propose to calibrate the tests using well founded statistical procedures for risk minimization : AIC criterion for the \code{fitAR} method,  bootstrap procedures for the \code{kernel} method and slope heuristics for the  \code{spectralproj} method. These procedures are implemented in the \slmf~function with the  \code{model\_selec = -1} argument, as detailed in the previous section.

\begin{figure}[h]
\begin{center}
\includegraphics[width=0.58\textwidth]{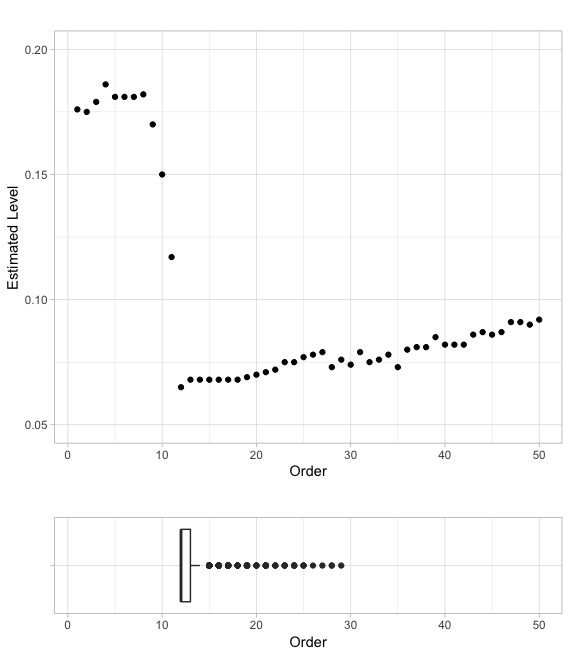}
\caption{Estimated level of the test according to the order of the fitted AR process on the residuals (top) and boxplot of the order selected by AIC, over $1000$ simulations. The data has been simulated according to Model \eqref{reglinsimu} with $\beta_1 = 3$ and $\beta_{2} = \beta_{3} = 0$, with $n= 1000$.}
\label{levels}
\end{center}
\end{figure}

Let us first illustrate the calibration problem with the AR12 process. For $T=1000$ simulations, we generate an error process of size $n$ under the null hypothesis: $H_{0}: \beta_{2} = \beta_{3} = 0$. Then we use the \code{fitAR} method of the \slmf~function with orders between $1$ and $50$ and we perform the model significance test. The procedure is repeated $1000$ times and we estimate the true level of the test by taking the average of the estimated levels on the $1000$ simulations for each order. The results are given on Figure~\ref{levels} for $n=1000$. 
A boxplot is also displayed to visualize the distribution of the order selected by the automatic criterion (AIC).

\subsection{Non-Seasonal errors}

We first study the case of non-Seasonal error processes. We simulate a $n$-error process according to the AR1, the Nonmixing or the Sysdyn processes. We simulate realizations of  the linear regression model \eqref{reglinsimu} under the null hypothesis: $H_{0}: \beta_{2} = \beta_{3} = 0$. We use the automatic  selection procedures for each method (\code{model\_selec = -1}).  The simulations are repeated $1000$ times in order to estimate the true level of the model significance for each test procedure. We simulate either small samples ($n=200$) or larger samples ($n=1000, 2000, 5000$). The results of this experiments are summarized in Table~\ref{tab:nonsea}.

\begin{table}[h]
\begin{center}
\begin{tabular}{|c|c|c|c|c|c|c|c|}
\hline
n & \backslashbox{Process}{Method} & Fisher test & fitAR & spectralproj & efromovich & kernel & hac \\
\hline
\multirow{3}*{200} & AR1 process & 0.465 & \textbf{0.097} & 0.14 & 0.135 & 0.149 & 0.108 \\
\cline{2-8}
& NonMixing & 0.298 & 0.082 & 0.103 & 0.096 & 0.125 & \textbf{0.064} \\
\cline{2-8}
& Sysdyn process & 0.385 & \textbf{0.105} & 0.118 & 0.124 & 0.162 & 0.12 \\
\hline
\multirow{3}*{1000} & AR1 process & 0.418 & 0.043 & \textbf{0.049} & \textbf{0.049} & 0.086 & \textbf{0.049} \\
\cline{2-8}
& NonMixing & 0.298 & 0.046 & \textbf{0.05} & 0.053 & 0.076 & 0.038 \\
\cline{2-8}
& Sysdyn process & 0.393 & \textbf{0.073} & 0.077 & 0.079 & 0.074 & 0.078 \\
\hline
\multirow{3}*{2000} & AR1 process & 0.454 & 0.071 & 0.078 & 0.075 & \textbf{0.067} & 0.071 \\
\cline{2-8}
& NonMixing & 0.313 & \textbf{0.051} & 0.053 & 0.057 & 0.067 & 0.047 \\
\cline{2-8}
& Sysdyn process & 0.355 & \textbf{0.063} & 0.064 & 0.066 & 0.069 & 0.073 \\
\hline
\multirow{3}*{5000} & AR1 process & 0.439 & 0.044 & \textbf{0.047} & \textbf{0.047} & \textbf{0.047} & 0.044 \\
\cline{2-8}
& NonMixing & 0.315 & 0.053 & 0.056 & 0.059 & 0.068 & \textbf{0.05} \\
\cline{2-8}
& Sysdyn process & 0.381 & 0.058 & 0.061 & \textbf{0.057} & 0.064 & 0.071 \\
\hline
\end{tabular} \\
\end{center}
\caption{Estimated levels for the non-seasonal processes.}
\label{tab:nonsea}
\end{table}


For $n$ large enough ($n \geq 1000$), all methods work well and the estimated level is around $0.05$. However, for small samples ($n = 200$), we observe that the \code{fitAR} and the \code{hac} methods show better performances than the others.
The \code{kernel} method is slightly less effective. With this method, we must choose the size of the bootstrap blocks as well as the number of blocks and the test results are really sensitive to these parameters. In these simulations, we have chosen $100$ blocks with a size of $n/2$. The results are expected to improve with a larger number of blocks.

Let us notice that for all methods and for all sample sizes, the estimated level is much better than if no correction is made (usual Fisher tests).

\subsection{Seasonal errors}

We now study the case of linear regression with seasonal errors. The experiment is exactly the same as before, except that we simulate AR12 or MA12 processes. The results of these experiments are summarized in Table~\ref{tab:sea}.

\begin{table}[h]
\begin{center}
\begin{tabular}{|c|c|c|c|c|c|c|c|}
\hline
n & \backslashbox{Process}{Method} & Fisher test & fitAR & spectralproj & efromovich & kernel & hac \\
\hline
\multirow{2}*{200} & AR12 process & 0.436 & 0.178 & 0.203 & 0.223 & 0.234 & \textbf{0.169} \\
\cline{2-8}
& MA12 process & 0.228 & \textbf{0.113} & \textbf{0.113} & 0.116 & 0.15 & 0.222 \\
\hline
\multirow{2}*{1000} & AR12 process & 0.468 & \textbf{0.068} & 0.183 & 0.181 & 0.124 & 0.179 \\
\cline{2-8}
& MA12 process & 0.209 & 0.064 & 0.066 & 0.069 & \textbf{0.063} & 0.18 \\
\hline
\multirow{2}*{2000} & AR12 process & 0.507 & \textbf{0.071} & 0.196 & 0.153 & 0.104 & 0.192 \\
\cline{2-8}
& MA12 process & 0.237 & 0.064 & 0.064 & \textbf{0.058} & 0.068 & 0.173 \\
\hline
\multirow{2}*{5000} & AR12 process & 0.47 & \textbf{0.062} & 0.183 & 0.1 & 0.091 & 0.171 \\
\cline{2-8}
& MA12 process & 0.242 & 0.044 & \textbf{0.048} & 0.043 & 0.057 & 0.147 \\
\hline
\end{tabular} \\
\end{center}
\caption{Estimated levels for the seasonal processes.}
\label{tab:sea}
\end{table}


We directly see  that the case of seasonal processes is more complicated than for the non-seasonal processes especially for the AR12 process. For small samples size, the estimated level is between $0.17$ and $0.24$, which is clearly too large. It is however much better than the estimated level of the usual Fisher test, which is around $0.45$.
The \code{fitAR} method is the best method here for the AR12 process, because for $n \geq 1000$, the estimated level is between $0.06$ and $0.07$.
For \code{efromovich} and \code{kernel} methods, a level less than $0.10$ is reached but for large samples only. The \code{spectralproj} and \code{hac} methods do not seem to work well for the AR12 process, although they remain much better than the usual Fisher tests (around $19\%$ of rejection instead of $45\%$). 

The case of the MA12 process seems easier to deal with. For $n$ large enough ($n \geq 1000$), the estimated level is between $0.04$ and $0.07$ whatever the method, except for \code{hac} (around $0.15$ for $n = 5000$). It is less effective for small sample size ($n=200$) with an estimated level around $0.115$ for \code{fitAR}, \code{spectralproj} and \code{efromovich} methods.

\subsection{I.I.D. errors}


To be complete, we consider in this subsection the case where the $\epsilon_{i}$'s are i.i.d., to see how the five automatic methods perform in that case. 

We simulate $n$ i.i.d. centered random variables according to the formula:
\[\epsilon_{i} = W_{i}^{2} - \frac{5}{4},\]
where $W$ follows a student distribution with $10$ degrees of freedom. Note that the distribution of the $\epsilon_{i}$'s is not symmetric and has no exponential moments. 

\begin{table}[h]
\begin{center}
\begin{tabular}{|c|c|c|c|c|c|c|c|}
\hline
n & \backslashbox{Process}{Method} & Fisher test & fitAR & spectralproj & efromovich & kernel & hac \\
\hline
150 & i.i.d. process & 0.053 & 0.068 & 0.078 & 0.061 & 0.124 & 0.063 \\
\hline
300 & i.i.d. process & 0.052 & 0.051 & 0.06 & 0.05 & 0.095 & 0.052 \\
\hline
500 & i.i.d. process & 0.047 & 0.049 & 0.053 & 0.049 & 0.082 & 0.056 \\
\hline
\end{tabular} \\
\end{center}
\caption{Estimated levels for the i.i.d. process}
\label{tab:i.i.d.}
\end{table}

Except for the kernel method, the estimated levels are close to $5 \%$ for $n$ large enough ($n \geq 300$).
It is slightly worse for small samples but it remains quite good for the methods \code{fitAR}, \code{efromovich} and \code{hac}.

As a general conclusion of Section $4$, we see that the \code{fitAR} method performs quite well in a wide variety of situations, and should therefore be used as soon as the user suspects that the error process can be modeled by a stationary short-memory process.

\section{Application to real data}
\label{sec:S5Data}

\subsection{Data CO2}
\label{sub:S5DataCO2}

\begin{figure}[h]
\begin{center}
\includegraphics[width=0.5\textwidth]{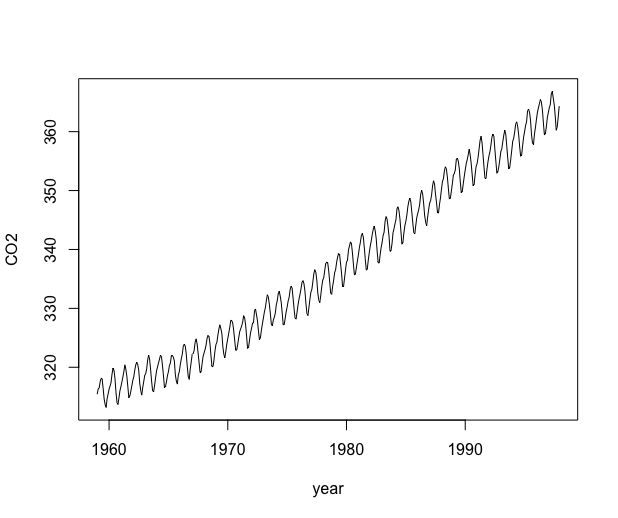}
\end{center}
\caption{CO2 rate as a function of time.}
\label{co2}
\end{figure}

Let us introduce the first dataset that we want to study. It concerns the well-known dataset "co2", available in the package \pkg{datasets} of \proglang{R}:
\begin{Schunk}
\begin{Sinput}
R> data("co2")
\end{Sinput}
\end{Schunk}
This dataset is provided by the observatory of Mona Loa (Hawaii). It contains average monthly measurements of CO2 (parts per million: ppmv) in the atmosphere of the Hawaiian coast. Surveys were produced monthly between $1959$ and $1998$, giving a total of $468$ measurements. The graph of the data is displayed in Figure~\ref{co2}. More information on this dataset is available in the \proglang{R} documentation.

We model the CO2 measurements with a time series. Typically, a time series can be decomposed into three parts: a trend $m$ and a seasonality $s$, which are deterministic components, and the errors $\varepsilon$, which constitute the random part of the model. The trend represents the overall behavior of the series and seasonality its periodic behavior. Formally, we have:
\[Y_{t} = m_{t} + s_{t} + \varepsilon_{t},\]
where $Y_t$ represents the CO2 rate at time $t$, with the usual constraints $s_{t} = s_{t+12}$ and $\sum_{t=1}^{12} s_{t} =0$.
The two deterministic components can be grouped into a matrix $X$  and the model can be rewritten into a linear regression model:
\[Y = X \beta + \varepsilon.\]
For this example, we fit a $3$-degree polynomial for the trend and a trigonometric polynomial with well-chosen frequencies for the seasonality. Here the time $t$ represents a month and $t$ goes from $1$ to $40$ by step of length $1/12$. Let us perform a linear regression to fit the trend and the seasonality on the CO2 time series, using the \code{lm} function:
\begin{Schunk}
\begin{Sinput}
R> y = as.vector(co2)
R> t = as.vector(time(co2)) - 1958
R> regtrigo = lm(y ~ t + I(t^2) + I(t^3) + sin(2*pi*t) + cos(2*pi*t)
+  	+ sin(4*pi*t) + cos(4*pi*t) + sin(6*pi*t) + cos(6*pi*t)
+  	+ sin(8*pi*t) + cos(8*pi*t))
\end{Sinput}
\end{Schunk}
We obtain the following output:
\begin{Schunk}
\begin{Sinput}
R> summary.lm(regtrigo)
\end{Sinput}
\begin{Soutput}
Call:
lm(formula = y ~ t + I(t^2) + I(t^3) + sin(2 * pi * t) + cos(2 * 
    pi * t) + sin(4 * pi * t) + cos(4 * pi * t) + sin(6 * pi * 
    t) + cos(6 * pi * t) + sin(8 * pi * t) + cos(8 * pi * t))

Residuals:
     Min       1Q   Median       3Q      Max 
-1.54750 -0.32688  0.00233  0.28100  1.50295 

Coefficients:
                  Estimate Std. Error  t value Pr(>|t|)    
(Intercept)      3.157e+02  1.118e-01 2823.332  < 2e-16 ***
t                3.194e-01  2.306e-02   13.847  < 2e-16 ***
I(t^2)           4.077e-02  1.293e-03   31.523  < 2e-16 ***
I(t^3)          -4.562e-04  2.080e-05  -21.930  < 2e-16 ***
sin(2 * pi * t)  2.751e+00  3.298e-02   83.426  < 2e-16 ***
cos(2 * pi * t) -3.960e-01  3.296e-02  -12.015  < 2e-16 ***
sin(4 * pi * t) -6.743e-01  3.296e-02  -20.459  < 2e-16 ***
cos(4 * pi * t)  3.785e-01  3.296e-02   11.484  < 2e-16 ***
sin(6 * pi * t) -1.042e-01  3.296e-02   -3.161  0.00168 ** 
cos(6 * pi * t) -4.389e-02  3.296e-02   -1.332  0.18362    
sin(8 * pi * t)  8.733e-02  3.296e-02    2.650  0.00833 ** 
cos(8 * pi * t)  2.559e-03  3.296e-02    0.078  0.93814    
---
Signif. codes:  0 '***' 0.001 '**' 0.01 '*' 0.05 '.' 0.1 ' ' 1

Residual standard error: 0.5041 on 456 degrees of freedom
Multiple R-squared:  0.9989,	Adjusted R-squared:  0.9989 
F-statistic: 3.738e+04 on 11 and 456 DF,  p-value: < 2.2e-16
\end{Soutput}
\end{Schunk}
We see in the summary that two variables have no significant effect on the CO2 rate. Next, we perform a backward selection method with a p-value threshold equal to $0.05$. This selects the following model:
\begin{Schunk}
\begin{Sinput}
R> regtrigo = lm(y ~ t + I(t^2) + I(t^3) + sin(2*pi*t) + cos(2*pi*t)
+  	+ sin(4*pi*t) + cos(4*pi*t) + sin(6*pi*t) + sin(8*pi*t))
\end{Sinput}
\end{Schunk}
with the corresponding summary
\begin{Schunk}
\begin{Sinput}
R> summary.lm(regtrigo)
\end{Sinput}
\begin{Soutput}
Call:
lm(formula = y ~ t + I(t^2) + I(t^3) + sin(2 * pi * t) + cos(2 * 
    pi * t) + sin(4 * pi * t) + cos(4 * pi * t) + sin(6 * pi * 
    t) + sin(8 * pi * t))

Residuals:
     Min       1Q   Median       3Q      Max 
-1.59287 -0.32364  0.00226  0.29884  1.50154 

Coefficients:
                  Estimate Std. Error  t value Pr(>|t|)    
(Intercept)      3.157e+02  1.118e-01 2824.174  < 2e-16 ***
t                3.196e-01  2.306e-02   13.861  < 2e-16 ***
I(t^2)           4.075e-02  1.293e-03   31.522  < 2e-16 ***
I(t^3)          -4.560e-04  2.080e-05  -21.927  < 2e-16 ***
sin(2 * pi * t)  2.751e+00  3.297e-02   83.446  < 2e-16 ***
cos(2 * pi * t) -3.960e-01  3.295e-02  -12.018  < 2e-16 ***
sin(4 * pi * t) -6.743e-01  3.295e-02  -20.464  < 2e-16 ***
cos(4 * pi * t)  3.785e-01  3.295e-02   11.487  < 2e-16 ***
sin(6 * pi * t) -1.042e-01  3.295e-02   -3.162  0.00167 ** 
sin(8 * pi * t)  8.734e-02  3.295e-02    2.651  0.00831 ** 
---
Signif. codes:  0 '***' 0.001 '**' 0.01 '*' 0.05 '.' 0.1 ' ' 1

Residual standard error: 0.504 on 458 degrees of freedom
Multiple R-squared:  0.9989,	Adjusted R-squared:  0.9989 
F-statistic: 4.57e+04 on 9 and 458 DF,  p-value: < 2.2e-16
\end{Soutput}
\end{Schunk}

\begin{figure}[h]
\begin{center}
\includegraphics[width=0.49\textwidth]{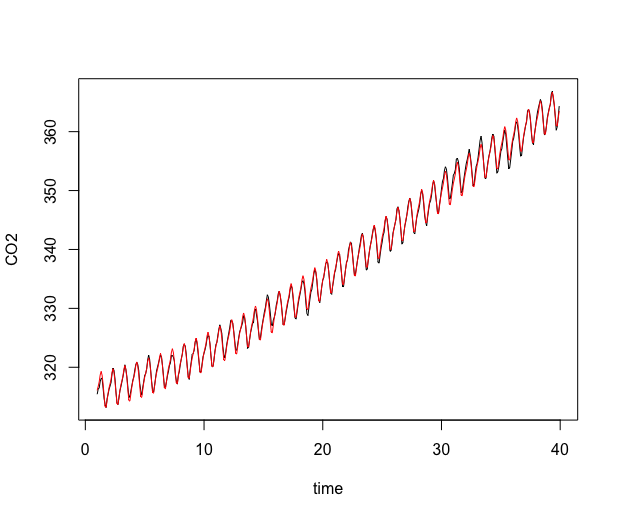}
\includegraphics[width=0.49\textwidth]{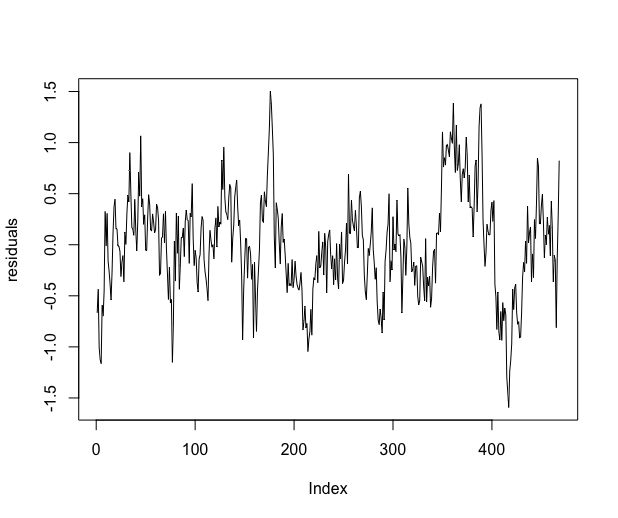}
\caption{CO2 adjustment (left) and residuals (right)}
\label{co2_ajust}
\end{center}
\end{figure}

\begin{figure}[h]
\begin{center}
\includegraphics[width=0.49\textwidth]{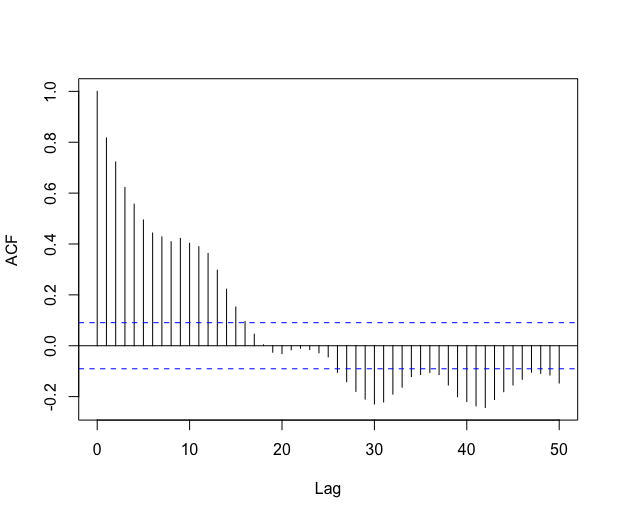}
\includegraphics[width=0.49\textwidth]{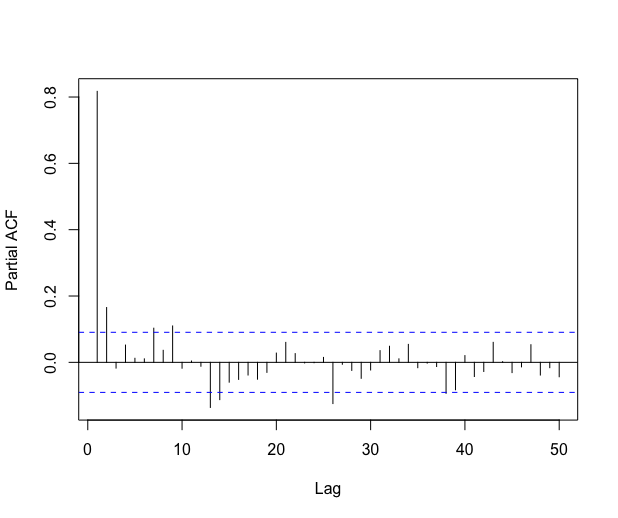}
\end{center}
\caption{Autocorrelation function (left) and partial autocorrelation function (right) of the residuals.}
\label{acf_resCO2}
\end{figure}

The sum of the estimated trend and estimated tendency is displayed on the left plot of Figure~\ref{co2_ajust}, and the residuals are displayed on the right plot.
The \code{lm} procedure assumes that the errors are independent, but if we look at the autocorrelation function of the residual process we clearly observe that the residuals are strongly correlated, see Figure~\ref{acf_resCO2}. Consequently,  the  \code{lm} procedure may be unreliable in this context.

\begin{figure}[h]
\begin{center}
\includegraphics[width=0.5\textwidth]{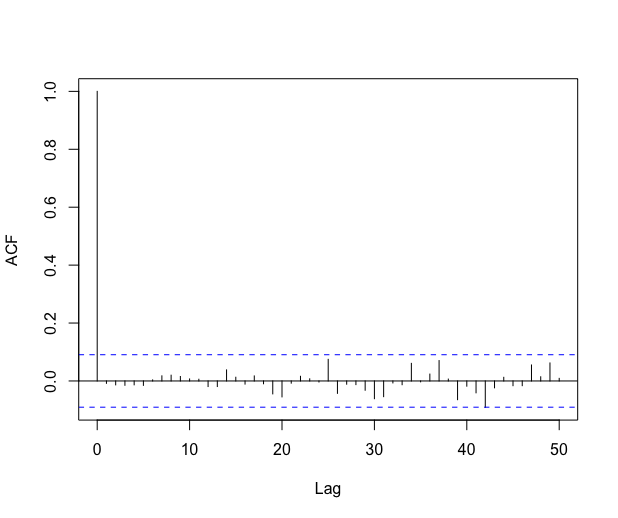}
\end{center}
\caption{Autocorrelation function of the residuals for the AR fitting.}
\label{res_fitARCO2}
\end{figure}

The autocorrelation function of the residuals decreases rather fast. Looking at the partial autocorrelation function, it seems reasonable to fit an AR process on the residuals. The automatic \code{fitAR} method selects an AR of order $14$ and the residuals look like a white noise, see Figure~\ref{res_fitARCO2}.

We now use the \slmf~function with the \code{fitAR} method  with the following complete model
\begin{Schunk}
\begin{Sinput}
R> regtrigo = slm(y ~ t + I(t^2) + I(t^3) + sin(2*pi*t) + cos(2*pi*t)
+  	+ sin(4*pi*t) + cos(4*pi*t) + sin(6*pi*t) + cos(6*pi*t)
+  	+ sin(8*pi*t) + cos(8*pi*t), method_cov_st = "fitAR",
+  	model_selec = -1)
\end{Sinput}
\end{Schunk}
Let us  display the summary of the procedure:
\begin{Schunk}
\begin{Sinput}
R> summary(regtrigo)
\end{Sinput}
\begin{Soutput}
Call:
"slm(formula = myformula, data = data, x = x, y = y)"

Residuals:
     Min       1Q   Median       3Q      Max 
-1.54750 -0.32688  0.00233  0.28100  1.50295 

Coefficients:
                  Estimate Std. Error z value Pr(>|z|)    
(Intercept)      3.157e+02  3.968e-01 795.646  < 2e-16 ***
t                3.194e-01  8.222e-02   3.884 0.000103 ***
I(t^2)           4.077e-02  4.619e-03   8.825  < 2e-16 ***
I(t^3)          -4.562e-04  7.430e-05  -6.140 8.23e-10 ***
sin(2 * pi * t)  2.751e+00  4.739e-02  58.054  < 2e-16 ***
cos(2 * pi * t) -3.960e-01  4.716e-02  -8.396  < 2e-16 ***
sin(4 * pi * t) -6.743e-01  2.051e-02 -32.875  < 2e-16 ***
cos(4 * pi * t)  3.785e-01  2.041e-02  18.548  < 2e-16 ***
sin(6 * pi * t) -1.042e-01  1.359e-02  -7.663 1.82e-14 ***
cos(6 * pi * t) -4.389e-02  1.359e-02  -3.228 0.001245 ** 
sin(8 * pi * t)  8.733e-02  1.246e-02   7.009 2.41e-12 ***
cos(8 * pi * t)  2.559e-03  1.252e-02   0.204 0.838038    
---
Signif. codes:  0 '***' 0.001 '**' 0.01 '*' 0.05 '.' 0.1 ' ' 1

Residual standard error: 0.5041
Multiple R-squared:  0.9989
chi2-statistic: 3.598e+04 on 11 DF,  p-value: < 2.2e-16
\end{Soutput}
\end{Schunk}
The last variable has no significant effect on the CO2. After performing a backward selection method with a p-value threshold  equal to $0.05$, we obtain the following model
\begin{Schunk}
\begin{Sinput}
R> regtrigo = slm(y ~ t + I(t^2) + I(t^3) + sin(2*pi*t) + cos(2*pi*t)
+  	+ sin(4*pi*t) + cos(4*pi*t) + sin(6*pi*t) + cos(6*pi*t)
+  	+ sin(8*pi*t), method_cov_st = "fitAR", model_selec = -1)
\end{Sinput}
\end{Schunk}
and the associated summary
\begin{Schunk}
\begin{Sinput}
R> summary(regtrigo)
\end{Sinput}
\begin{Soutput}
Call:
"slm(formula = myformula, data = data, x = x, y = y)"

Residuals:
     Min       1Q   Median       3Q      Max 
-1.54877 -0.32432  0.00187  0.28069  1.50168 

Coefficients:
                  Estimate Std. Error z value Pr(>|z|)    
(Intercept)      3.157e+02  3.969e-01 795.627  < 2e-16 ***
t                3.194e-01  8.223e-02   3.884 0.000103 ***
I(t^2)           4.077e-02  4.619e-03   8.825  < 2e-16 ***
I(t^3)          -4.562e-04  7.430e-05  -6.140 8.23e-10 ***
sin(2 * pi * t)  2.751e+00  4.738e-02  58.061  < 2e-16 ***
cos(2 * pi * t) -3.960e-01  4.716e-02  -8.397  < 2e-16 ***
sin(4 * pi * t) -6.743e-01  2.051e-02 -32.874  < 2e-16 ***
cos(4 * pi * t)  3.785e-01  2.041e-02  18.547  < 2e-16 ***
sin(6 * pi * t) -1.042e-01  1.359e-02  -7.664 1.80e-14 ***
cos(6 * pi * t) -4.389e-02  1.359e-02  -3.229 0.001244 ** 
sin(8 * pi * t)  8.733e-02  1.248e-02   6.998 2.60e-12 ***
---
Signif. codes:  0 '***' 0.001 '**' 0.01 '*' 0.05 '.' 0.1 ' ' 1

Residual standard error: 0.5036
Multiple R-squared:  0.9989
chi2-statistic: 3.596e+04 on 10 DF,  p-value: < 2.2e-16
\end{Soutput}
\end{Schunk}
There is a clear difference between the two backward procedures: \slmf~keeps the variable $\cos(6 \pi x)$, while \code{lm} does not. Given the obvious dependency of the error process, we recommend using \slmf~instead of \code{lm} in this context.

\subsection{PM2.5 Data of Shanghai}
\label{sub:S5DataChine}

This dataset comes from a study about fine particle pollution in five Chinese cities. The data are available on the following website \url{https://archive.ics.uci.edu/ml/datasets/PM2.5+Data+of+Five+Chinese+Cities#}.
We are interested here by the city of Shanghai. We study the regression of PM2.5 pollution in Xuhui District by other  measurements of pollution in neighboring districts and also by meteorological variables. The dataset contains  hourly observations between January 2010 and December 2015. More precisely it contains $52584$ records of $17$ variables: date, time of measurement, pollution and meteorological variables. More information on these data is available in the paper of~\cite{liang2016pm2}.

We remove the lines that contain NA observations and we then extract the first $5000$ observations. For simplicity, we will only consider pollution variables and weather variables. We start the study with the following $10$ variables:
\begin{itemize}
\item[-] PM\_Xuhui: PM2.5 concentration in the Xuhui district ($ug/m^{3}$)
\item[-] PM\_Jingan: PM2.5 concentration in the Jing'an district ($ug/m^{3}$)
\item[-] PM\_US.Post: PM2.5 concentration in the U.S diplomatic post ($ug/m^{3}$)
\item[-] DEWP: Dew Point (Celsius Degree)
\item[-] TEMP: Temperature (Celsius Degree)
\item[-] HUMI: Humidity ($\%$)
\item[-] PRES: Pressure (hPa)
\item[-] Iws: Cumulated wind speed ($m/s$)
\item[-] precipitation: hourly precipitation (mm)
\item[-] Iprec: Cumulated precipitation (mm)
\end{itemize}
\begin{Schunk}
\begin{Sinput}
R> shan = read.csv("ShanghaiPM20100101_20151231.csv", header = TRUE, 
+  	sep = ",")
R> shan = na.omit(shan)
R> shan_complete = shan[1:5000,c(7,8,9,10,11,12,13,15,16,17)]
R> shan_complete[1:5,]
\end{Sinput}
\begin{Soutput}
      PM_Jingan PM_US.Post PM_Xuhui DEWP  HUMI PRES TEMP Iws
26305        66         70       71   -5 69.00 1023    0  60
26306        67         76       72   -5 69.00 1023    0  62
26308        73         78       74   -4 74.41 1023    0  65
26309        75         77       77   -4 80.04 1023   -1  68
26310        73         78       80   -4 80.04 1023   -1  70
      precipitation Iprec
26305             0     0
26306             0     0
26308             0     0
26309             0     0
26310             0     0
\end{Soutput}
\end{Schunk}

The aim is to study the concentration of particles in Xuhui District according to the other variables. We first fit a linear regression with the \code{lm}  function:
\begin{Schunk}
\begin{Sinput}
R> reglm = lm(shan_complete$PM_Xuhui ~ . ,data = shan_complete)
R> summary.lm(reglm)
\end{Sinput}
\begin{Soutput}
Call:
lm(formula = shan_complete$PM_Xuhui ~ ., data = shan_complete)

Residuals:
     Min       1Q   Median       3Q      Max 
-132.139   -4.256   -0.195    4.279  176.450 

Coefficients:
                Estimate Std. Error t value Pr(>|t|)    
(Intercept)   -54.859483  40.975948  -1.339 0.180690    
PM_Jingan       0.596490   0.014024  42.533  < 2e-16 ***
PM_US.Post      0.375636   0.015492  24.246  < 2e-16 ***
DEWP           -1.038941   0.170144  -6.106 1.10e-09 ***
HUMI            0.291713   0.045799   6.369 2.07e-10 ***
PRES            0.025287   0.038915   0.650 0.515852    
TEMP            1.305543   0.168754   7.736 1.23e-14 ***
Iws            -0.007650   0.002027  -3.774 0.000163 ***
precipitation   0.462885   0.132139   3.503 0.000464 ***
Iprec          -0.125456   0.039025  -3.215 0.001314 ** 
---
Signif. codes:  0 '***' 0.001 '**' 0.01 '*' 0.05 '.' 0.1 ' ' 1

Residual standard error: 10.68 on 4990 degrees of freedom
Multiple R-squared:  0.9409,	Adjusted R-squared:  0.9408 
F-statistic:  8828 on 9 and 4990 DF,  p-value: < 2.2e-16
\end{Soutput}
\end{Schunk}
The variable PRES has no significant effect on the PM\_Xuhui variable. We then perform a backward selection procedure, which leads to select $9$ significant variables:
\begin{Schunk}
\begin{Sinput}
R> shan_lm = shan[1:5000,c(7,8,9,10,11,13,15,16,17)]
R> reglm = lm(shan_lm$PM_Xuhui ~ . ,data = shan_lm)
R> summary.lm(reglm)
\end{Sinput}
\begin{Soutput}
Call:
lm(formula = shan_lm$PM_Xuhui ~ ., data = shan_lm)

Residuals:
     Min       1Q   Median       3Q      Max 
-132.122   -4.265   -0.168    4.283  176.560 

Coefficients:
                Estimate Std. Error t value Pr(>|t|)    
(Intercept)   -28.365506   4.077590  -6.956 3.94e-12 ***
PM_Jingan       0.595564   0.013951  42.690  < 2e-16 ***
PM_US.Post      0.376486   0.015436  24.390  < 2e-16 ***
DEWP           -1.029188   0.169471  -6.073 1.35e-09 ***
HUMI            0.285759   0.044870   6.369 2.08e-10 ***
TEMP            1.275880   0.162453   7.854 4.90e-15 ***
Iws            -0.007734   0.002023  -3.824 0.000133 ***
precipitation   0.462137   0.132127   3.498 0.000473 ***
Iprec          -0.127162   0.038934  -3.266 0.001098 ** 
---
Signif. codes:  0 '***' 0.001 '**' 0.01 '*' 0.05 '.' 0.1 ' ' 1

Residual standard error: 10.68 on 4991 degrees of freedom
Multiple R-squared:  0.9409,	Adjusted R-squared:  0.9408 
F-statistic:  9933 on 8 and 4991 DF,  p-value: < 2.2e-16
\end{Soutput}
\end{Schunk}

\begin{figure}[h]
\begin{center}
\includegraphics[width=0.49\textwidth]{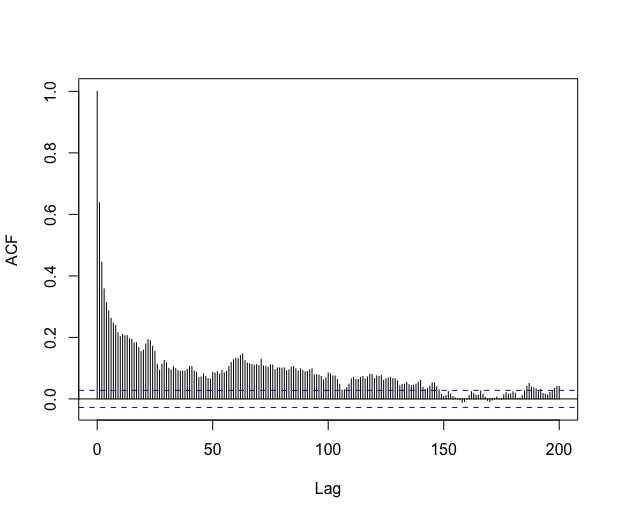}
\includegraphics[width=0.49\textwidth]{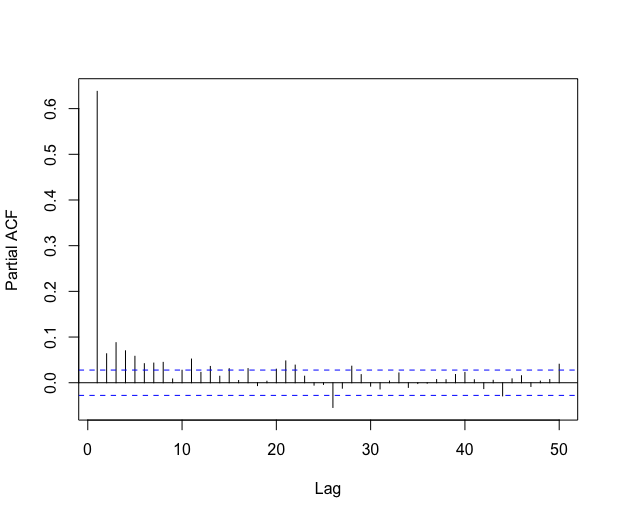}
\end{center}
\caption{Autocorrelation function (left) and partial autocorrelation function (right) of the residuals.}
\label{acf_pacf_shan}
\end{figure}

The autocorrelation of the residual process shows that the errors are clearly not i.i.d., see Figure~\ref{acf_pacf_shan}. We thus suspect the \code{lm} procedure to be unreliable in this context.

\begin{figure}[h]
\begin{center}
\includegraphics[width=0.5\textwidth]{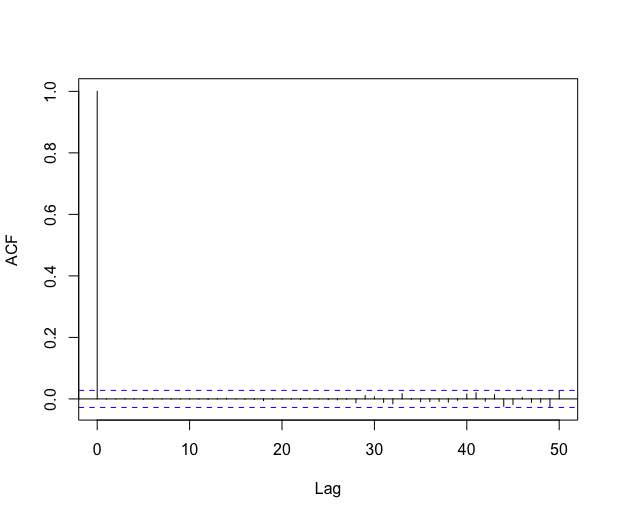}
\end{center}
\caption{Autocorrelation function of the residuals for the AR fitting.}
\label{res_shan_AR28}
\end{figure}

The autocorrelation function  decreases pretty fast, and the partial autocorrelation function suggests that fitting an AR process on the residuals should be an appropriate method in this case. The automatic \code{fitAR} method of \slmf~selects an AR process of order $28$. The residuals of this AR fitting look like a white noise, as shown in Figure~\ref{res_shan_AR28}.
Consequently, we propose to perform a linear regression with \slmf~function, using the fitAR method on the complete model
\begin{Schunk}
\begin{Sinput}
R> regslm = slm(shan_complete$PM_Xuhui ~ . ,data = shan_complete,
+  	method_cov_st = "fitAR", model_selec = -1)
R> summary(regslm)
\end{Sinput}
\begin{Soutput}
Call:
"slm(formula = myformula, data = data, x = x, y = y)"

Residuals:
     Min       1Q   Median       3Q      Max 
-132.139   -4.256   -0.195    4.279  176.450 

Coefficients:
                Estimate Std. Error z value Pr(>|z|)    
(Intercept)   -54.859483 143.268399  -0.383 0.701783    
PM_Jingan       0.596490   0.028467  20.953  < 2e-16 ***
PM_US.Post      0.375636   0.030869  12.169  < 2e-16 ***
DEWP           -1.038941   0.335909  -3.093 0.001982 ** 
HUMI            0.291713   0.093122   3.133 0.001733 ** 
PRES            0.025287   0.137533   0.184 0.854123    
TEMP            1.305543   0.340999   3.829 0.000129 ***
Iws            -0.007650   0.005698  -1.343 0.179399    
precipitation   0.462885   0.125641   3.684 0.000229 ***
Iprec          -0.125456   0.064652  -1.940 0.052323 .  
---
Signif. codes:  0 '***' 0.001 '**' 0.01 '*' 0.05 '.' 0.1 ' ' 1

Residual standard error: 10.68
Multiple R-squared:  0.9409
chi2-statistic:  8383 on 9 DF,  p-value: < 2.2e-16
\end{Soutput}
\end{Schunk}
Note that the variables show  globally larger p-values than with the \code{lm} procedure, and more variables have no significant effect than with \code{lm}. After performing a backward selection we obtain the following results
\begin{Schunk}
\begin{Sinput}
R> shan_slm = shan[1:5000,c(7,8,9,10,11,13)]
R> regslm = slm(shan_slm$PM_Xuhui ~ . , data = shan_slm, 
+  	method_cov_st = "fitAR", model_selec = -1)
R> summary(regslm)
\end{Sinput}
\begin{Soutput}
Call:
"slm(formula = myformula, data = data, x = x, y = y)"

Residuals:
     Min       1Q   Median       3Q      Max 
-132.263   -4.341   -0.192    4.315  176.501 

Coefficients:
             Estimate Std. Error z value Pr(>|z|)    
(Intercept) -29.44924    8.38036  -3.514 0.000441 ***
PM_Jingan     0.60063    0.02911  20.636  < 2e-16 ***
PM_US.Post    0.37552    0.03172  11.840  < 2e-16 ***
DEWP         -1.05252    0.34131  -3.084 0.002044 ** 
HUMI          0.28890    0.09191   3.143 0.001671 ** 
TEMP          1.30069    0.32435   4.010 6.07e-05 ***
---
Signif. codes:  0 '***' 0.001 '**' 0.01 '*' 0.05 '.' 0.1 ' ' 1

Residual standard error: 10.71
Multiple R-squared:  0.9406
chi2-statistic:  8247 on 5 DF,  p-value: < 2.2e-16
\end{Soutput}
\end{Schunk}
The backward selection with \slmf~only keeps $5$ variables.

\section*{Acknowledgements}
The authors are grateful to Anne Philippe and Aymeric Stamm for valuable discussions.

\bibliography{slm_bib}

\end{document}